\documentclass[12pt]{article}
\usepackage{psfig}
\begin{document}
\newcommand{\beq}{\begin{equation}}
\newcommand{\eeq}{\end{equation}}

\begin{center}
{\Large \bf $^3$P$_2$--$^3$F$_2$ Pairing in Dense
Neutron Matter:\\
The Spectrum of Solutions}
\end{center}

\vskip 1 cm

\centerline{\large M. V. Zverev$^1$, J. W. Clark$^2$, and V. A.
Khodel$^{1,2}$ }

\vskip 1 cm

\begin{center}
{\it $^1$Russian Research Center Kurchatov Institute,\\
Moscow, 123182, Russia \\
 $^2$McDonnell Center for the Space
Sciences and Department of Physics,\\
Washington University, St. Louis, MO 63130-4899
}
\end{center}

\vskip 0.5 cm \hrule \vskip 0.5 cm

\begin{abstract}
The $^3$P$_2$--$^3$F$_2$ pairing model is generally considered to provide
an adequate description of the superfluid states of neutron matter
at densities some 2-3 times that of saturated symmetrical nuclear
matter.  The problem of solving the system of BCS gap equations
expressing the $^3$P$_2$--$^3$F$_2$ model is attacked with the aid of
the separation approach.  This method, developed originally for
quantitative study of S-wave pairing in the presence of strong
short-range repulsions, serves effectively to reduce the coupled,
singular, nonlinear BCS integral equations to a set of coupled
algebraic equations.  For the first time, sufficient precision becomes
accessible to resolve small energy splittings between the different
pairing states.  Adopting a perturbative strategy, we are able to
identify and characterize the full repertoire of real solutions of the
$^3$P$_2$--$^3$F$_2$ pairing model, in the limiting regime of small
tensor-coupling strength.  The P--F channel coupling is seen to lift
the striking parametric degeneracies revealed by a earlier separation
treatment of the pure, uncoupled $^3$P$_2$ pairing problem.  Remarkably,
incisive and robust results are obtained solely on the basis of analytic
arguments.  Unlike the traditional Ginzburg-Landau approach, the analysis
is not restricted to the immediate vicinity of the critical temperature,
but is equally reliable at zero temperature.  Interesting connections
and contrasts are drawn between triplet pairing in dense neutron matter
and triplet pairing in liquid $^3$He.
\end{abstract}

\vskip 0.5 cm \hrule \vskip 0.5 cm

\section{Introduction}

It is widely accepted that triplet pairing between constituent
spin-1/2 fermions gives rise to superfluid phases in liquid $^3$He
at millikelvin temperatures and in the neutronic component of the
quantum fluid interior of a neutron star at temperatures in the
hundred keV range and below.  In both cases, the density is so high
that the familiar singlet S-wave gap is quenched by the dominant
effect of the short-range repulsion in that channel.  Instead, pairing
is favored in a channel with an odd orbital momentum $L=1$,
and therefore in the triplet spin state.  The pairing mechanisms
active in these two examples produce interesting distinctions
between them.  In superfluid $^3$He, pairing is triggered by spin
fluctuations \cite{vol}, and the B-state (or B-phase) with total angular
momentum $J=0$ occupies most of the superfluid phase diagram.  However,
the spin-fluctuation mechanism induces relatively tiny energy
splittings between states with different $J$ values.  Consequently,
very close to the critical temperature $T_c$ there exists a
phase transition from the B-phase to the A-phase, which involves
a combination of $J=1$ and $J=2$ pairing channels.

The situation in neutron matter is rather different.  Referring to
the experimental data on the energy dependence of the $nn$ scattering
phase shifts \cite{arndt}, one may infer that central forces between a
neutron pair are quite weak, and that pairing must be due predominantly
to the spin-orbit force.  Based on empirical analyses of spin-orbit
splitting in finite nuclei, the latter force is expected to be
insensitive to polarization and correlation effects.  The
spin-orbit pairing mechanism implies that the spin ${\bf S}$ and
orbital momentum ${\bf L}$ of the pair cease to be conserved
separately, and pairing in the $J=2$ channel dominates
\cite{hof,tam,takt,ost,tt,kkc2,kkc3}.

In both Fermi quantum liquids, the effective in-medium interaction
contains a component that mixes two-body states with orbital angular
momenta $L\pm~2$.  Specifically, pairing channels with $L=1$ and $L=3$
are coupled by the magnetic dipole force in the case of liquid $^3$He
and by the tensor force arising from pion exchange, in neutron matter.
However, the effect is of minuscule importance in the $^3$He problem,
since the magnetic dipole component is smaller in magnitude than the
dominant central part of the interaction by a factor $10^{-7}$~\cite{vol}.
By contrast, the magnitude of the dominant spin-orbit force in neutron
matter is only a few times larger than the tensor component, so the effects
of the latter cannot be neglected.  (More specifically, the parameter
measuring the strength of the tensor force relative to the spin-orbit
force varies around $0.3$ in the density interval $\rho_0<\rho<3\rho_0$
\cite{arg,kkc3} if in-vacuum interaction constants are adopted,
$\rho_0$ being the saturation density of symmetrical nuclear matter.)
Moreover, pion exchange is responsible for the most powerful
fluctuations in neutron matter.  Accordingly, it is imperative to
give careful attention to\\
$^3$P$_2$--$^3$F$_2$ channel coupling
in the quantitative description of pairing in neutron matter.

Reliable prediction of the phase diagram of triplet pairing over an
extensive temperature range has proven to be extraordinarily difficult.
The traditional tool for elucidation of the phase diagram in liquid $^3$He
and other systems manifesting superfluidity or superconductivity has
been the Ginzburg-Landau functional approach.  But since Ginzburg-Landau
theory is valid only near the critical temperature, it has little relevance
to neutron stars, which cool down to temperatures one or two orders of
magnitude below $T_c$ by a thousand years after their birth in a supernova
event.  There is of course the alternative of brute-force iterative solution
of the gap equation, which has been widely used in quantitative studies
of singlet S-wave pairing \cite{chen}.  However, this strategy is
generally afflicted with slow convergence and uncertain accuracy both
for S-wave interactions containing strong short-range repulsions and
for pairing in higher angular momentum states, where one must deal
with a multitude of coupled nonlinear singular integral equations.
The limitations of standard iterative approaches become particularly
serious when one seeks to construct the superfluid phase diagram
of the system, which is sensitive to tiny energy splittings between
the different solutions of the BCS pairing problem.

Explication of the complete superfluid phase diagram of neutron matter
has two facets.  First, one must identify and characterize the set
of admissible solutions of the BCS gap equations arising in the
$^3$P$_2$--$^3$F$_2$ pairing problem, i.e., one must find the
``spectrum of solutions.''  Second, one must determine the relative
stability of the different solutions under variation of density,
temperature, and other relevant parameters, so as to uncover the
possible phase transitions and map out the actual phase diagram.

The purpose of this paper is to present a detailed account of
the substantial progress that has been made on the first of these
tasks through application of the separation method developed
in ref.~\cite{kkc1,kkc2} for robust and accurate solution of
BCS gap equations.  (The reader may consult ref.~\cite{vak1}
for a comprehensive review of the separation approach.)
Many of the results obtained here have been exposed in condensed
form in earlier works \cite{kkc2,kcz}; it is our intent here to
offer a more complete justification of these findings, and to
provide an enhanced understanding of their wider implications.

In the separation method, the BCS system is recast so as to isolate the
major, logarithmically divergent contributions to the pairing effect and
treat them separately from the remaining features of the problem,
which are very insensitive to the presence of the gap and its
particular value.  This method, which in essence reduces the problem to
solution of system of algebraic equations, is equally reliable and
precise in the limiting regimes $T\to T_c$ and $T\to 0$, as well
as in between.  In concrete calculations, we may accommodate and
exploit the fact, based on the experimental $P$-scattering phases
\cite{arndt}, that the central components of the in-vacuum $nn$ interaction
nearly compensate each other. We {\it assume} -- quite plausibly -- that
this feature is maintained by the {\it effective} $nn$ interaction within
neutron matter.  We furthermore assume -- with less confidence --
that the smallness of the parameter characterizing the importance
of the tensor force relative to the spin-orbit force is maintained in
dense neutron matter.  It must be acknowledged that medium modification
may be more significant in the case of the tensor force than it is
for the spin-orbit component, especially in the vicinity of the
phase transition leading to pion condensation \cite{mig,akmal}.
Investigation of this issue calls for a concerted effort and
will be pursued elsewhere.

The current study of the $^3$P$_2$--$^3$F$_2$ model is exhaustive in its
assembly of the set of solutions whose structural expression involves
only real numbers.  The implementation of this program requires no
numerical computations; remarkably, everything can be done analytically.
In principle, the same method can be readily applied to the determination
of complex solutions.  However, this more complicated problem
inevitably entails some computer work.

The paper is organized as follows:  In Sec.~2, the coupled-channel
BCS formalism is stated, the separation method is applied to the gap
equations, and further analysis establishes the explicit equations
of an incisive perturbative treatment of the $^3$P$_2$--$^3$F$_2$
pairing model.  In Sec.~3, the parametric degeneracy inherent
in the unperturbed $^3$P$_2$ problem is lifted, and all real
solutions of the perturbed model are found.  Section 4 furnishes a
convenient and succinct catalog of these solutions.  Finally,
Sec.~5 is devoted to a discussion of potential vulnerabilities
of our approach to the $^3$P$_2$--$^3$F$_2$ problem in neutron matter,
as well as informative connections with aspects of triplet pairing
in liquid $^3$He.

\section{The Basic Set of Gap Equations}

Fixing the spin and isospin of the pairing state at $S=1$ and
$T=1$ (triplet-triplet), we start with the partial-wave decomposition
\begin{equation}
\Delta_{\alpha\beta}({\bf p}) =\sum_{J,L,M}\Delta_L^{JM}(p)
\end{equation}
of the $2\times 2$ gap matrix in terms of the spin-angle matrices
\begin{equation}
   \left(G_{LJ}^M({\bf n})\right)_{\alpha\beta}=\sum_{M_SM_L}
   C^{1M_S}_{{1\over 2}{1\over 2}\alpha\beta}C^{JM}_{1LM_SM_L}
   Y_{LM_L}({\bf n})
\end{equation}
and the multi-channel BCS gap equations
\begin{eqnarray}
     \Delta_L^{JM}(p)
 &=& \sum_{L'L_1J_1M_1}(-1)^{1+{L-L'\over 2}}\int \int
     \langle p| V_{LL'}^J |p_1 \rangle
     S^{JMJ_1M_1}_{L'L_1}({\bf n}_1) \nonumber  \\
&\qquad&~ \times
  {\tanh \left(E({\bf p}_1) / 2T\right)\over 2 E({\bf p}_1)}
  \Delta_{L_1}^{J_1M_1}(p_1) p^2_1 dp_1d{\bf n}_1
\label{bc}
\end{eqnarray}
that embrace the formal problem to be solved \cite{ost,tt,kkc2,kkc3}.
The latter equations contain the spin trace
\begin{equation}
   S_{LL_1}^{JMJ_1M_1} ({\bf n})
 = {\rm Tr} \left[\left(G_L^{JM}({\bf n})\right)^*
   G_{L_1}^{J_1M_1}({\bf n})\right]
\end{equation}
and the interaction matrix elements
$\langle p| V_{LL'}^J|p_1 \rangle$ appearing in
the partial-wave expansion
\begin{equation}
   V({\bf  p},{\bf p}_1)
 = \sum_{LL'JM}(-1)^{{L-L'\over 2}}
   \langle p| V_{LL'}^J|p_1 \rangle G_{LJ}^M({\bf n})
   \left(G_{L'J}^M({\bf n}_1)\right)^*
\end{equation}
of the block of Feynman diagrams irreducible in the particle-particle
channel.  The quasiparticle energy
\begin{eqnarray}
        E({\bf p})
    &=& \sqrt{\xi^2(p)+D^2({\bf p})}\nonumber \\
    &=& \left[\xi^2(p)+{1\over 2}\sum_{LJML_1J_1M_1}
        \left(\Delta_L^{JM}(p)\right)^* \Delta_{L_1}^{J_1M_1}(p)
        S^{JMJ_1M_1}_{LL_1}({\bf n})\right]^{1\over 2}
\label{qen}
\end{eqnarray}
is assembled from the single-particle spectrum $\xi(p)$ of the
normal Fermi liquid, measured relative to the chemical
potential $\mu$ and often parametrized with an effective mass
$M^*$, together with the gap components $\Delta_L^{JM}(p)$.
We note that $E({\bf p})$ takes on an angular dependence by
virtue of the spin trace $S_{LL_1}^{JMJ_1M_1}({\bf n})$,
which in principle greatly complicates the task of solving
the system (\ref{bc}).  However, this angular dependence
only comes into play near the Fermi surface; hence it can
be ignored in those integral contributions to the r.h.s.\ of
Eq.~(\ref{bc}) in which the region around the Fermi surface is
suppressed \cite{kkc2,kkc3}.  Consequently, in practice the gap
equations approximately decouple in the variables $L'$, $L_1$, and
$J_1$, by virtue of the orthogonality property
\begin{equation}
   \int S_{LL_1}^{JMJ_1M_1}({\bf n})d{\bf n}
   =\delta_{LL_1}\delta_{JJ_1}\delta_{MM_1} \,.
\end{equation}
Treatment of the exact, coupled equations (\ref{bc}) is therefore not
as difficult as it appears at first sight (though still not at all
trivial).

The angle dependence of the function $D({\bf p})$ makes
it awkward to speak of {\it the} energy gap.  Therefore it is
conventional to introduce the quantity
\begin{equation}
   \Delta_F = \left[{\overline{D^2}}(k_F)\right]^{1/2}
\label{egap}
\end{equation}
as a representative measure of the pairing gap in the quasiparticle
spectrum, where the overbar signifies an angle average.

As we have argued in the introduction, the spin-orbit component of the
neutron-neutron interaction exerts a strong influence on pair formation
in dense neutron matter, favoring the condensation of pairs in the
$^3$P$_2$ state.  Accepting this widely held view, the simplifying feature
just revealed implies that contributions to triplet pairing from
nondiagonal terms with $L',\, L_1\neq 1$ or $J_1\neq 2$ on the
r.h.s.\ of Eq.~(1) can be evaluated within perturbation theory.
Defining $v_F\equiv\langle p_F| V_{11}^2|p_F \rangle$, the relevant
coupling parameter is $\eta=-\langle p_F| V_{13}^2|p_F \rangle/v_F$.
The evaluation is carried out in terms of the set of ``principal'' gap
amplitudes $\Delta_1^{2M}(p)$, with $M$ running from $-2$ to 2.  A
further simplification ensues from time-reversal invariance, which
implies that only three of these five quantities can be independent,
say $\Delta_1^{2M}(p)$ with $M=0\,,1\,,2$.

Focusing on the nondiagonal contributions to the r.h.s.\ of Eq.~(1),
we observe that two of them are of leading significance.
The first contains the integral of the product
$ V_{31}^2 S^{2M2M_1}_{31}\Delta^{2M_1}_1$, and
the second contains the integral of the product
$ V_{11}^2S^{2M2M_1}_{13}\Delta^{2M_1}_3$.  Restricting attention
to these contributions, we arrive at the $^3$P$_2$--$^3$F$_2$ pairing
problem, which has been studied both analytically and numerically
in earlier work \cite{takt,ost,bal,kkc3}.  The list of participating
states appears to be exhausted upon addition of the $^3$P$_0$ and
$^3$P$_1$ pairing channels, which are deemed to be of lesser
importance; their role will be examined later in this paper.

Adopting the $^3$P$_2$--$^3$F$_2$ pairing model as circumscribed above,
the BCS system (\ref{bc})
takes the explicit form
\begin{eqnarray}
       \Delta_1^{2M}(p)
 &+& \sum_{M_1} \int \int
     \langle p| V_{11}^2 |p_1 \rangle S^{2M2M_1}_{11}({\bf n}_1)
     {\tanh \left(E({\bf p}_1) / 2T\right)\over 2 E({\bf p}_1)}
     \Delta_1^{2M_1}(p_1) p^2_1 dp_1d{\bf n}_1 \nonumber\\
 &=& \sum_{M_1}\int \int
     \langle p| V_{13}^2 |p_1 \rangle S^{2M2M_1}_{31}({\bf n}_1)
     {\tanh \left(E_0({\bf p}_1) / 2T\right)\over 2 E_0({\bf p}_1)}
     \Delta_1^{2M_1}(p_1) p^2_1 dp_1d{\bf n}_1\nonumber\\
 &+&~ \sum_{M_1}\int \int
      \langle p| V_{11}^2 |p_1 \rangle S^{2M2M_1}_{13}({\bf n}_1)
      {\tanh \left(E_0({\bf p}_1) / 2T\right)\over 2 E_0({\bf p}_1)}
      \Delta_3^{2M_1}(p_1) p^2_1 dp_1d{\bf n}_1\,,\nonumber\\
      \Delta_3^{2M}(p)
 &=& \sum_{M_1} \int \int
     \langle p| V_{31}^2 |p_1 \rangle S^{2M2M_1}_{11}({\bf n}_1)
     {\tanh \left(E_0({\bf p}_1) / 2T\right)\over 2 E_0({\bf p}_1)}
     \Delta_1^{2M_1}(p_1) p^2_1 dp_1d{\bf n}_1   \,,\nonumber\\
\label{bc1}
\end{eqnarray}
which will now be subjected to analysis and solution.  In writing
the r.h.s.\ of this equation, we have replaced the quasiparticle
energy $E({\bf p};\eta)$ by\\
$E_0({\bf p};\eta=0)= \left[\xi^2(p)+D^2_0({\bf p})\right]^{1/2}$,
where $D_0({\bf p})$
is the gap function of the much-studied $^3$P$_2$ pairing model
in which the tensor coupling between $F$ and $P$ states is ignored.
The substitution $E \rightarrow E_0$ in the integrals on the right
is justified by the small relative size of the quantities
$\langle p| V_{13}^2 |p_1 \rangle$, $\Delta_3^{2M_1}(p_1)$,
and $\langle p| V_{31}^2 |p_1 \rangle$.

A conspicuous feature of the $^3$P$_2$ pairing model is the high
parametric degeneracy of the spectrum of its solutions.  This
property has been investigated in detail in refs.~\cite{kkc2,kkc3}
in terms of the two ratios
 $\lambda_1=D^{21}_1\sqrt{6}/D^{20}_1
            =-D^{2,-1}_1\sqrt{6}/D^{20}_1$ and
 $\lambda_2=D^{22}_1\sqrt{6}/D^{20}_1
            =D^{2,-2}_1\sqrt{6}/D^{20}_1$.
The degeneracy is reflected
in the existence of a set of curves $\lambda_1(\lambda_2)$
in the plane $(\lambda_1,\lambda_2)$, upon which all the
BCS equations of the $^3$P$_2$ problem are satisfied.  As we shall
see, this degeneracy is essentially lifted in the $^3$P$_2$--$^3$F$_2$
pairing model where $\eta\neq 0$.  A finite set  of points
$(\lambda_1,\lambda_2)$, depending somehow on the $\eta$ value,
replaces the set of solution curves $\lambda_1(\lambda_2)$ of the
$^3$P$_2$ model.

It is the objective of this article to identify the different solutions
of the system (\ref{bc1}) and to establish their structure in the
realistic case of small $\eta$.  The analysis is aided by the fact that
the parameters $\lambda_1=f_1(\eta)$ and $\lambda_2= f_2(\eta)$
are {\it continuous functions} of the coupling constant $\eta$.
This property implies that the number of solutions of the
$^3$P$_2$--$^3$F$_2$
pairing problem as well as their structure remains the same no
matter how small $\eta$ is.  Consequently, implementation of our
program reduces to determination of the functions $f_1$ and $f_2$
at $\eta=0$, i.e., $\lambda_1(\eta=0)$ and $\lambda_2(\eta=0)$.
It then becomes apparent that the quasiparticle energy $E(\eta)$ may
be replaced by $E_0$ on the left-hand-sides of Eqs.~(\ref{bc1}) as well
as on the right, since taking into account the difference $E(\eta)$
and $E_0$ within (\ref{bc1}) cannot, by itself, lift the parametric
degeneracy.  This conclusion is confirmed in the numerical calculations
we have performed.

The nondiagonal integrals on the r.h.s.\ of Eqs.~(\ref{bc1}) are
rapidly convergent, with the overwhelming contributions coming from
momenta adjacent to the Fermi surface.  This feature greatly expedites
application of the perturbation strategy.  For $E({\bf p})$ significantly
in excess of the energy gap $\Delta_F$ of Eq.~(\ref{egap}), the energies
$E({\bf p})$ and $|\xi(p)|$ are coincident to high precision, such that
the angular integration in Eq.~(\ref{bc1}) yields a null result.  Thus,
when treating the nondiagonal contributions it is sufficient
to know the minor gap components $\Delta_3^{2M}(p)$  at the
point $p=p_F$, which may be efficiently evaluated in terms of the
coefficients $D^{2M}_1\equiv \Delta^{2M}_1(p_F)$ (with $M=0\,,1\,,2$).
In this process, we retain, on the r.h.s.\ of the last of Eqs.~(\ref{bc1}),
only the dominant contribution containing a large logarithmic factor
$L=\ln (\epsilon_F/\Delta_F)$, where $\epsilon_F$ is the Fermi energy.
This factor is angle-independent; therefore the respective
angular integral is freely evaluated, giving rise to the simple connection
\begin{equation}
     \Delta^{2M}_3(p=p_F)
  = -L\langle p_F| V_{13}^2|p_F \rangle
     D^{2M}_1=\eta v_F L D^{2M}_1\simeq \eta D^{2M}_1 \ .
\label{prop1}
\end{equation}
In obtaining this
relation we have employed the equality $1=v_FL$, which holds when one
keeps only logarithmic contributions.  Analogous linear relations are
obtained for the other minor components $\Delta_L^{JM}$ of the gap function
(notably $\Delta_1^{00}$ and $\Delta_1^{1M}$).

Insertion of the result (\ref{prop1}) into the first of Eqs.~(\ref{bc1})
leads to the closed system of equations
$$
     \Delta_1^{2M}(p) +\sum_{M_1}
      \int\int \langle p| V_{11}^2 |p_1 \rangle
      S^{2M2M_1}_{11}({\bf n}_1)
      {\tanh \left(E_0({\bf p}_1) / 2T\right)\over 2 E_0({\bf p}_1)}
      \Delta_1^{2M_1}(p_1) p^2_1 dp_1d{\bf n}_1
$$
\begin{eqnarray}
    =&& \sum_{M_1}\int \int
        \langle p| V_{13}^2 |p_1 \rangle
        \left[S^{2M2M_1}_{31}({\bf n}_1)
              +S^{2M2M_1}_{13}({\bf n}_1)\right]\nonumber\\
     &\times& {\tanh \left(E_0({\bf p}_1) / 2T\right)\over 2 E_0({\bf p}_1)}
              \Delta_1^{2M_1}(p_1) p^2_1 dp_1d{\bf n}_1
\label{bcf}
\end{eqnarray}
for finding the set of three gap functions $\Delta^{2M}_1(p)$ with
$M=0,1,2$.

At this point we invoke the separation method and assert the
decomposition \cite{kkc2,kkc3}
\begin{equation}
   \Delta^{2M}_1(p)\equiv D^{2M}_1\chi(p)
\end{equation}
of the gap component into a ``universal'' shape factor $\chi(p)$
that is {\it independent} of the magnetic quantum number $M$ and
a numerical coefficient $D^{2M}_1$ that embodies the dependence on
$M$.  The function $\chi(p)$, normalized by $\chi(p_F)=1$, is the
solution of a {\it linear} integral equation.  As argued in
refs.~\cite{kkc2,kkc3}, this decomposition holds to high accuracy
in the problem domain under consideration.
Accordingly, our problem reduces to the determination of the three
key coefficients $D^{2M}_1$, which obey a set of coupled algebraic
equations obtained by setting $p=p_F$ in Eqs.~(\ref{bcf}).  With
$M=0$, 1, and 2, these equations read
\begin{eqnarray}
     D^{2M}_1
 &+&v_F\sum_{M_1} D^{2M_1}_1
      \int\int\phi(p) {\tanh{\left( E_0({\bf p})/ 2T \right)}
      \over 2E_0({\bf p})} S^{2M2M_1}_{11}({\bf n})
      \chi(p)p^2 dp d{\bf n} \nonumber \\
 &=& \eta v_F \sum_{M_1}D_1^{2M_1} \int
     \left[S^{2M2M_1}_{31}({\bf n})+S^{2M2M_1}_{13}({\bf n})\right]
     K_0({\bf n})d{\bf n}  \,,
\label{cutr}
\end{eqnarray}
with $\phi(p)\equiv\langle p| V^2_{11}|p_F \rangle/v_F$, and
\begin{equation}
   K_0({\bf n})=\int\langle p| V_{13}^2 |p_1 \rangle
   {\tanh{\left(E_0({\bf p})/2T\right)}\over 2E_0({\bf p})}\chi(p)
    p^2 dp  \, .
\label{ko}
\end{equation}
It should be remarked that the integral (\ref{ko}) contains a
constant coming from regions lying far from the Fermi surface, but
this constant does not contribute to the
angular integration on the r.h.s.\ of Eq.~(\ref{cutr}).

In the present work, the search for solutions will be confined to those
with real coefficients $D^{2M}_1$.  The structure of the phases having
complex coefficients $D^{2M}_1$ can be explored and established along the
same lines, although the calculations become much more cumbersome.  To
streamline the task of finding solutions, it is helpful to rewrite
Eqs.~(\ref{cutr}) in terms of
the ratios $\lambda_1$ and $\lambda_2$.
This step ensures
coincidence between the left-hand sides of these equations and
their counterparts in the model of pure $^3$P$_2$ pairing solved
in ref.\cite{kkc2}.  Substitution of the explicit form of
$S^{2M2M_1}_{11}({\bf n})$ into Eqs.~(\ref{cutr}), followed by
straightforward algebra, gives a system of three equations
\begin{eqnarray}
     \lambda_2+ v_F\left[\lambda_2(J_0+J_5) -\lambda_1 J_1 -J_3\right]
 &=& \eta v_F r_2\, ,\nonumber \\
     \lambda_1+ v_F\left[-(\lambda_2+1)J_1+\lambda_1(J_0+4J_5+2J_3)/4\right]
 &=& \eta v_F  r_1\, ,\nonumber\\
     1+ v_F\left[-(\lambda_2 J_3+\lambda_1 J_1)/3 +J_5\right]
 &=&\eta v_F  r_0 \,,
\label{sprl}
\end{eqnarray}
for these ratios and the gap value $\Delta_F$.  Here, as before~\cite{kkc2},
\begin{equation}
    J_i
  = \int\!\!\int f_i(\vartheta,\varphi)
    \varphi(p){\tanh{E_0({\bf p})/ 2T}\over 2E_0({\bf p})}
    \chi(p) {p^2dpd{\bf n}\over 4\pi}
\label{intk}
\end{equation}
with $f_0=1-3z^2$, $f_1=3xz/2$, $f_3= 3(2x^2+z^2-1)/2$, and\\
$f_5=(1+3z^2)/2$ and $z=\cos\vartheta$, $x=\sin\vartheta\cos\varphi$,
and $y=\sin\vartheta\sin\varphi$.  Among the integrals $J_i$
($i=1,\cdots 5$), only $J_5$ contains a singular principal term going like
$\ln(\epsilon_F/\Delta_F)$.  All the other $J_i$ converge close to the Fermi
surface, where $E_0({\bf p})=\left[\xi^2(p)+D^2_0({\bf n})\right]^{1/2}$
with
\begin{eqnarray}
       D^2_0({\bf n})
  &=& {\Delta^2_F \over 2[1+(\lambda_1^2+\lambda_2^2)/3]}
      \biggl[ 1+3\cos^2\theta+\lambda^2_2\sin^2\theta+
      {\lambda^2_1\over 2}(1+\cos^2\theta) \nonumber \\
 &\quad&~ -2\lambda_1(1+\lambda_2)\cos\theta\sin\theta\cos\varphi
          +{1\over 2}(\lambda^2_1-4\lambda_2)\sin^2\theta\cos2\phi \biggr] \,.
\label{do}
\end{eqnarray}
In terms of the coefficients
$\lambda_1,\lambda_2$, the right-hand sides of Eqs.~(\ref{sprl}) read
\begin{eqnarray}
      r_2
  &=& \lambda_2s_{22}+\lambda_1s_{21}+\sqrt{6}s_{20}\ , \nonumber \\
      r_1
  &=& \lambda_2s_{12}+\lambda_1s_{11}+\sqrt{6}s_{10}\, \nonumber \\
      r_0
  &=& {\lambda_2\over \sqrt{6}}s_{02}
      +{\lambda_1\over \sqrt{6}} s_{01} +s_{00} \, ,
\label{rhsr}
\end{eqnarray}
where
\begin{eqnarray}
      s_{M2}
 &=& \int \left[ S^{2M22}_{31}({\bf n})+S^{2M22}_{13}({\bf n})
     +S^{2M2,-2}_{31}({\bf n})+S^{2M2,-2}_{13}({\bf n}) \right]
     K_0({\bf n})d{\bf n}  \ , \nonumber \\
     s_{M1}
 &=& \int \left[ S^{2M21}_{31}({\bf n})+S^{2M21}_{13}({\bf n})
     -S^{2M2,-1}_{31}({\bf n})-S^{2M2,-1}_{13}({\bf n}) \right]
     K_0({\bf n})d{\bf n}  \ , \nonumber \\
     s_{M0}
 &=& \int \left[ S^{2M20}_{31}({\bf n})+S^{2M20}_{13}({\bf n})\right]
     K_0({\bf n})d{\bf n}  \ .
\label{intr}
\end{eqnarray}

\section{Real Solutions of the $^3$P$_2$--$^3$F$_2$ Problem}

Equations~(\ref{sprl}) have three familiar {\it one-component} solutions
with definite magnetic quantum numbers $M=0$, 1, and 2.  To uncover
the structure and the spectrum of the {\it multicomponent}, mixed-$M$
solutions of the perturbed problem, a two-step transformation is applied
to the system (\ref{sprl}).  The integral $J_5$ introduces the gap value
$\Delta_F$ into the description, but it is irrelevant to the phase
structure.  As a first step, we combine Eqs.~(\ref{sprl}) so as to
eliminate terms involving $J_5$ from the first pair and, at the same time,
reduce the number of the $J_i$ integrals in each equation to two. The
resulting equations are
\begin{eqnarray}
(\lambda_2+1)[  3\lambda_1(\lambda_2+1)J_0 -
           2(\lambda^2_1-2\lambda_2^2+6) J_1] &=&\eta B_1
   \, ,\nonumber \\
(\lambda_2+1)[ (\lambda_1^2-4\lambda_2)J_1 +
                    \lambda_1(\lambda_2+1)J_3] &=&\eta B_2 \  ,
\label{spp2}
\end{eqnarray}
with
\begin{eqnarray}
     B_1 &=& 2\lambda_1(2\lambda_2+3)r_2-4(\lambda^2_2-3)
             r_1 -6\lambda_1(\lambda_2+2)r_0\  , \nonumber\\
     B_2 &=& -\lambda_1 r_2+4\lambda_2 r_1-3\lambda_1\lambda_2  r_0 \  .
\label{bi}
\end{eqnarray}

Two cases are to be distinguished.  The first corresponds to $\lambda_2=-1$,
a particular solution of the pure $^3$P$_2$ pairing problem \cite{kkc2}.
In this case, the left-hand sides of Eqs.~(\ref{spp2}) vanish identically,
and hence so must their right-hand sides, leading to the single
restriction
\begin{equation}
      \lambda_1 r_2(\lambda_1;-1)+4 r_1(\lambda_1;-1)-3
      \lambda_1 r_0(\lambda_1;-1) =0
\label{rel1} \,.
\end{equation}
As will be seen, this condition is satisfied at {\it any} $\lambda_1$.
Therefore the particular solution $\lambda_2=-1$ found in the pure
$^3$P$_2$ problem survives intact when the $^3$F$_2$ coupling
is switched on.

Let us now assume $\lambda_2\neq  -1$ and proceed to the second step.
Following ref.~\cite{kkc2}, we perform a rotation
\begin{equation}
   \left\{\begin{array}{c}
       z = t\cos\beta + u \sin\beta \,, \\
       x = -t\sin\beta + u \cos\beta \,,
\end{array}\right.
\label{rot}
\end{equation}
with the objective of removing the integral $J_1$ from Eqs.~(\ref{spp2}).
To this end, the parameter $\zeta=\tan\beta$ is taken as a root of
the quadratic equation
\begin{equation}
\label{first}
\zeta^2 + \frac{3-\lambda_2}{\lambda_2}\zeta - 1 = 0 \,.
\end{equation}
The integrals $J_i$ transform according to
\begin{eqnarray}
         J_0
  &\to & J_0(\cos^2\beta-{1\over 2}\sin^2\beta)- J_3\sin^2\beta \,, \nonumber\\
         J_1
  &\to & -{3\over 4}J_0 \sin^2\beta+J_3
          \Bigl(\cos^2\beta+{1\over 2} \sin^2\beta\Bigr)\,,\nonumber\\
         J_1
  &\to& \Bigl({3\over 4}J_0+{1\over 2}J_3\Bigr)\sin\beta\cos\beta   \,.
\end{eqnarray}
Substitution of the transformed integrals into Eqs.~(\ref{spp2})
yields
\begin{eqnarray}
  (\lambda_2+1)[A_1J_0+A_2J_3]&=&\eta B_1 \,, \nonumber \\
  (\lambda_2+1)[A_1J_0+A_2J_3]&=&-2 \eta B_2  \,,
\label{eqaa}
\end{eqnarray}
with
\begin{eqnarray}
      A_1
  &=& {3\over 2}\lambda_1(1+\lambda_2)(2-\zeta^2) -{3\over 2}
      (\lambda_1^2-2\lambda_2^2+6)\zeta \,, \nonumber \\
      A_2
  &=& -3\lambda_1(1+\lambda_2)\zeta^2
      -(\lambda_1^2 -2\lambda_2^2+6)\zeta \,.
\end{eqnarray}
(Some details of this step are provided in refs.~\cite{kkc2,vak1}.)

The left-hand members of the two equations (\ref{eqaa}) are seen to be
identical.  In fact, the universalities of pure $^3$P$_2$ pairing
revealed in ref.~\cite{kkc2} stem from this key property.  The solutions
of the restricted problem derived from Eqs.~(\ref{eqaa}) at
$\eta\equiv 0$
fall into two groups composed of states that are degenerate in energy.
This remarkable feature is independent of temperature, density, and
details of the in-medium interaction.  There is an upper (i.e.,
higher-energy) group consisting of states whose angle-dependent order
parameters have nodes and a lower group without nodes
(cf.\ ref.~\cite{rich}).  In addition to the energy degeneracies,
the multicomponent pairing solutions, which obey the relation
\begin{equation}
   (\lambda^2_1-2\lambda^2_2-6\lambda_2) (\lambda^2_1+2-2\lambda_2)
   = 0 \, ,
\label{specc}
\end{equation}
manifest a parametric degeneracy with respect to the
coefficient ratios
$\lambda_1$ and $\lambda_2$, as they in general define {\it curves}
rather than {\it points} in the $(\lambda_1,\lambda_2)$ plane.

Among solutions of the $^3$P$_2$--$^3$F$_2$ pairing model there exist some
for which $B_1$ and $B_2$ in Eq.~(\ref{eqaa}) vanish simultaneously
at a certain set of parameters $\lambda_1$ and $\lambda_2$
satisfying Eq.~(\ref{specc}).  Indeed, consider the set of values
\begin{equation}
   \lambda_1=0;\quad \lambda_2=\pm 1,\pm 3  \,,
\label{two}
\end{equation}
which determine two-component solutions of the pure $^3$P$_2$ pairing problem
\cite{kkc2}.  In these cases, it may be observed from Eqs.~(\ref{bi})
and (\ref{rhsr}) that
\begin{eqnarray}
    &&   B_1(\lambda_1=0,\lambda_2)
    \sim B_2(\lambda_1=0,\lambda_2)
    \sim r_1(\lambda_1=0,\lambda_2)\nonumber\\
    &=& \lambda_2 s_{12} (\lambda_1=0,\lambda_2)
     +\sqrt{6}s_{10} (\lambda_1=0,\lambda_2) \,,
\end{eqnarray}
where
   $s_{12}(\lambda_1=0,\lambda_2)$ and $s_{10}(\lambda_1=0,\lambda_2)$
are defined
by Eq.~(\ref{intr}).  As seen from Eq.~(\ref{do}), the quantities
$D_0(\lambda_1=0,\lambda_2;{\bf n})$ and
$K_0(\lambda_1=0,\lambda_2;{\bf n})$ are even functions of
$\cos\varphi$ \cite{ost,kkc2}, while both
$S^{2122}_{31}({\bf n})+S^{2122}_{13}({\bf n})
+S^{212,-2}_{31}({\bf n})+S^{212,-2}_{13}({\bf n})$
and $S^{2120}_{31}({\bf n})+S^{2120}_{13}({\bf n})$
appear to be linear in $\cos\varphi$ (for details, refer to the appendix).
As a result, both the matrix elements
$s_{12}(\lambda_1=0,\lambda_2)$ and $s_{10}(\lambda_1=0,\lambda_2)$
vanish identically when the integration
over $\varphi$ is performed.  Hence both of the quantities $B_1$ and $B_2$
turn out to be zero, and we find that Eq.~(31) defines a discrete set
of valid two-component solutions for the\\
$^3$P$_2$--$^3$F$_2$ pairing
problem as well as for the pure $^3$P$_2$ case.  No other two-component
solutions of the former problem have been found in numerical calculations.

Continuing our exploration, suppose that $B_1$ and $B_2$ do not vanish.
Equating the right-hand members of Eqs.~(\ref{eqaa}) in
the $\eta\to 0$ limit, we then obtain an {\it additional} relation between
the parameters $\lambda_1(\eta=0)$ and $\lambda_2(\eta=0)$ similar
to (\ref{rel1}), namely
\begin{equation}
    \lambda_1 r_2(\lambda_1,\lambda_2)
  -(\lambda_2-3) r_1(\lambda_1,\lambda_2)
  -3\lambda_1  r_0(\lambda_1,\lambda_2)=0 \,,
\label{rela}
\end{equation}
the quantities $r_M$ being defined by Eq.~(\ref{rhsr}).  Inserting
the explicit expressions for the $r_M$, this auxiliary condition can
be recast as
\begin{eqnarray}
    && G(\lambda_1,\lambda_2)\nonumber\\
    &=& \int K_0(x,y,z;\lambda_1,\lambda_2)
        \Psi(x,y,z;\lambda_2)\delta(1-x^2-y^2-z^2)
        \frac{dx\,dy\,dz}{2\pi}=0 \, ,
\label{g12}
\end{eqnarray}
where $\Psi(x,z;\lambda_1,\lambda_2)\equiv
\lambda_1 R_2(x,z)-(\lambda_2-3) R_1(x,z)-3\lambda_1  R_0(x,z)$
and the quantities $R_k$ are given in the appendix.  Substituting
for the $R_k$, we obtain
\begin{eqnarray}
 &&    \Psi(x,z;\lambda_1,\lambda_2)\nonumber\\
 &=& 14\lambda_1 \left[ 5(\lambda_2+3)z^4
     -10\lambda_2x^4 +15(\lambda_2-3)x^2z^2
     +6(\lambda_2+1)(x^2-z^2) \right] \nonumber \\
 &\quad&~ +70\left[(\lambda_2-3)(\lambda_2+3)+2\lambda_1^2
             \right] xz^3 \nonumber \\
 &\quad&~ +70\left[2\lambda_2(\lambda_2-3)-2\lambda_1^2
             \right] x^3z -96 (\lambda_2-3)(\lambda_2+1)xz \, .
\label{psi12}
\end{eqnarray}

The relation (\ref{g12}) supplements the spectral condition (\ref{specc}).
As a direct consequence, the strong parametric degeneracy inherent in
pure $^3$P$_2$ pairing is lifted in the case of $^3$P$_2$--$^3$F$_2$ pairing.
With the exception of the straight-line solution $\lambda_2 = 1$
noted above, the solutions of the problem are now represented by a
set of {\it isolated} points in the $(\lambda_1,\lambda_2)$ plane.

The system formed by Eqs.~(\ref{specc}) and (\ref{rela}) is amenable
to analytic solution.  We begin the search for solutions of Eq.~(\ref{g12})
with the particular solution ($\lambda_1$, $\lambda_2=-1$) of the pure
$^3$P$_2$ pairing problem.  In this case,
\begin{eqnarray}
         \Psi(x,y,z;\lambda_1,\lambda_2=-1)
  &\sim& z^4+x^4-6x^2z^2\nonumber\\
  &=& 8z^4-3(1-y^2)^2 - 4(1-y^2)(2z^2-1+y^2) \,,
\end{eqnarray}
while, as seen from (\ref{do}), the gap function
$D^2_0(\lambda_2=-1;{\bf n})$ depends only on the single variable $y$:
\begin{equation}
D^2_0(x,y,z;\lambda_1,\lambda_2=-1)\sim{3\over 2}(x^2+z^2)={3\over 2}(1-y^2)\,.
\end{equation}
Since the integrals
\begin{equation}
   \int\limits_{-\sqrt{1-y^2}}^{\sqrt{1-y^2}}
   \frac{8z^4-3(1-y^2)^2}{\sqrt{1-y^2-z^2}} \,dz \qquad {\rm and} \qquad
   \int\limits_{-\sqrt{1-y^2}}^{\sqrt{1-y^2}}
   \frac{2z^2-1+y^2}{\sqrt{1-y^2-z^2}} \,dz
\end{equation}
both vanish, the integral $G(\lambda_1,\lambda_2=-1)$ also vanishes
at any $\lambda_1$.  Accordingly, the degenerate solution
$\lambda_2=-1$ of the uncoupled $^3$P$_2$ pairing problem survives
when the $^3$F$_2$ channel is involved.

In search of other solutions, we again make use of the transformation
(\ref{rot}), applying it now to the whole integrand of Eq.~(\ref{g12}).
As found in ref.~\cite{kkc2}, the gap function $D_0$ reduces to a
function of the single variable $t$ under such a transformation;
hence the same property holds for the factor $K_0$ in Eq.~(\ref{g12}),
which is defined as a functional of $D_0$ by Eq.~(\ref{ko}).
To ascertain how the factor $\Psi$ is transformed, let us write down the
results of the transformation for simple terms entering this function.
In implementing the transformation we omit odd-power terms $u$ and
$u^3$, which do not contribute to the integral (\ref{g12}).  One obtains
\begin{eqnarray}
        x^4
  &\to& t^4 \sin^4\beta + u^4 \cos^4 \beta
        + 6t^2u^2\sin^2\beta\cos^2\beta\,,\nonumber\\
        z^4
  &\to& t^4 \cos^4\beta + u^4 \sin^4 \beta
        + 6 t^2 u^2\sin^2\beta\cos^2\beta\,,\nonumber\\
        x^2z^2
  &\to& t^4 \sin^2\beta\cos^2\beta
        + u^4 \sin^2\beta\cos^2\beta
        + t^2u^2(\sin^4\beta+\cos^4\beta
        -4\sin^2\beta\cos^2\beta)\,,\nonumber\\
        xz^3
  &\to& u^4\sin^3\beta\cos\beta - t^4\cos^3\beta\sin\beta
        + 3t^2u^2 (\cos^3\beta\sin\beta
        - \sin^3\beta\cos\beta)\,,\nonumber\\
        x^3z
  &\to& u^4\cos^3\beta\sin\beta - t^4\sin^3\beta\cos\beta
        - 3t^2u^2 (\cos^3\beta\sin\beta
        - \sin^3\beta\cos\beta)\,,\nonumber\\
        x^2-z^2
  &\to& (t^2-u^2) (\cos^2\beta-\sin^2\beta) \,.
\label{xz2}
\end{eqnarray}
After inserting these relations into the formula (\ref{psi12}) for
$\Psi$, simple algebra yields
\begin{equation}
     \Psi(t,u;\lambda_1,\lambda_2)
   = (1+\zeta^2)^{-2}
     \left[ U u^4 + Tt^4 + V(u^2-t^2) + Wu^2t^2\right]\,,
\end{equation}
where
\begin{eqnarray}
     U(\lambda_1,\lambda_2)
 &=& 70\biggl\{ \lambda_1(\lambda_2+3)\zeta^4
     + \left[(\lambda_2-3)(\lambda_2+3)
     +2\lambda_1^2\right] \zeta^3 \nonumber \\
 &\quad&~+ 3\lambda_1 (\lambda_2-3) \zeta^2
         + \left[2\lambda_2(\lambda_2-3)-2\lambda_1^2\right]
            \zeta -2\lambda_1\lambda_2 \biggr\} \,,
\end{eqnarray}
\begin{eqnarray}
     T(\lambda_1,\lambda_2)
 &=& 70\biggl\{ -2\lambda_1\lambda_2\zeta^4
     -[2\lambda_2(\lambda_2-3)-2\lambda_1^2]\zeta^3
     + 3\lambda_1 (\lambda_2-3) \zeta^2 \nonumber \\
 &\quad&~- [(\lambda_2-3)(\lambda_2+3)+2\lambda_1^2] \zeta
         + \lambda_1(\lambda_2+3)\biggr\} \,,
\end{eqnarray}
\begin{eqnarray}
      V(\lambda_1,\lambda_2)
  &=& 24 \biggl[ 7\lambda_1(\lambda_2+1)(1+\zeta^2)(1-\zeta^2) \nonumber\\
  &\quad&~-8(\lambda_2-3)(\lambda_2+1)(1+\zeta^2) \zeta\biggr] \,,
\end{eqnarray}
\begin{eqnarray}
     W(\lambda_1,\lambda_2)
 &=& 210\biggl\{ 2\lambda_1(\lambda_2+3)
     \zeta^2 - 4\lambda_1\lambda_2\zeta^2
     +\lambda_1 (\lambda_2-3) (\zeta^4+1-4\zeta^2) \nonumber \\
 &\quad&~+\left[(\lambda_2-3)(\lambda_2+3) +2\lambda_1^2\right]
         (\zeta-\zeta^3) \nonumber\\
 &\quad&~+ \left[2\lambda_2(\lambda_2-3)
         - 2\lambda_1^2\right](\zeta^3-\zeta) \biggr\} \ .
\end{eqnarray}
These results can be simplified slightly by employing the connection\\
$\lambda_1 (\zeta^2-1) = (\lambda_2-3) \zeta$, and
one finally arrives at
\begin{eqnarray}
      U(\lambda_1,\lambda_2)
  &=& 70 \biggl[ \lambda_1(\lambda_2+3)\zeta^4
      +5\lambda_1 (\lambda_2-3)\zeta^2 -2\lambda_1\lambda_2\nonumber\\
  &\quad&~ + (\lambda_2-3)(\lambda_2+3)\zeta^3
           + 2\lambda_2(\lambda_2-3)\zeta \biggr] \,,
\end{eqnarray}
\begin{eqnarray}
          T(\lambda_1,\lambda_2)
   &=&    70 \biggl[  \lambda_1(\lambda_2+3)
          +5\lambda_1 (\lambda_2-3)\zeta^2
          -2\lambda_1\lambda_2 \zeta^4 \nonumber\\
   &\quad&~ - (\lambda_2-3)(\lambda_2+3)\zeta
           - 2\lambda_2(\lambda_2-3)\zeta^3 \biggr]\,,
\end{eqnarray}
\begin{eqnarray}
      V(\lambda_1,\lambda_2)
  &=& 180\lambda_1(\lambda_2+1)(1-\zeta^4) \,,\nonumber\\
      W(\lambda_1,\lambda_2)
  &=& 420 \lambda_1(\lambda_2-3) (\zeta^4-6\zeta^2+1) \,.
\end{eqnarray}
The auxiliary integrals
\begin{eqnarray}
      I_0
  &=& \int\limits_{-\sqrt{1-t^2}}^{\sqrt{1-t^2}}
      \frac{du}{\sqrt{1-t^2-u^2}} = \pi \ ,  \nonumber\\
      I_2
  &=& \int\limits_{-\sqrt{1-t^2}}^{\sqrt{1-t^2}}
      \frac{u^2 du}{\sqrt{1-t^2-u^2}} = \frac{\pi(1-t^2)}{2} \,,
\end{eqnarray}
and
\begin{equation}
   I_4 = \int\limits_{-\sqrt{1-t^2}}^{\sqrt{1-t^2}}
         \frac{u^4 du}{\sqrt{1-t^2-u^2}} = \frac{3\pi(1-t^2)^2}{8}
\end{equation}
are helpful in completing the evaluation of $G(\lambda_1,\lambda_2)$
by integration over the new variables.  The integrals $I_0$, $I_2$,
and $I_4$ are related by
\begin{equation}
  I_4 = {3\over 4} (1-t^2) I_2 = {3\over 8} (1-t^2)^2 I_0 \ , \qquad
  I_2 = {1\over 2} (1-t^2) I_0 \, .
\end{equation}
Using these connections, one can easily verify that integration of the
combinations $8u^4-3(1-t^2)^2$ and $2u^2-1+t^2$
over $u$ gives zero.  It follows then that if we make the following
replacements
\begin{equation}
    u^2 \to {1\over 2}(1-t^2), \quad
    u^4 \to {3\over 8} (1-t^2)^2 \
\end{equation}
in Eq.~(\ref{psi12}), the new function
\begin{eqnarray}
     \Psi'(t,\lambda_1,\lambda_2)
  &=& \frac{1}{8(1+\zeta^2)^2}
      \Bigl\{ \left[ 3 U(\lambda_1,\lambda_2)
      + 8 T(\lambda_1,\lambda_2)
      - 4 W(\lambda_1,\lambda_2)\right] t^4 \nonumber \\
&\quad&~ - \left[ 6U(\lambda_1,\lambda_2)
        + 12 V(\lambda_1,\lambda_2)
        -4W(\lambda_1,\lambda_2)\right] t^2 \nonumber \\
&\quad&~ + 3 U(\lambda_1,\lambda_2)
        + 4V(\lambda_1,\lambda_2) \Bigr\}
\end{eqnarray}
will guarantee the same result as given by $\Psi$ upon integration of
Eq.~(\ref{g12}).  For this integral to vanish identically, with
the function $K_0(t)$ regarded as arbitrary, the coefficients
of all powers of $t$ in the function $\Psi'$ must be zero:
\begin{equation}
  \left\{ \begin{array}{c}
         3U(\lambda_1,\lambda_2)+8T(\lambda_1,\lambda_2)
         -4W(\lambda_1,\lambda_2) = 0\,, \\
         3U(\lambda_1,\lambda_2) + 6V(\lambda_1,\lambda_2)
         - 2W(\lambda_1,\lambda_2) = 0\,, \\
         3U(\lambda_1,\lambda_2) + 4V(\lambda_1,\lambda_2) = 0 \, .
\end{array}
\right.
\end{equation}
This system reduces to the chain of equations
\begin{equation}
%\label{set}
    T(\lambda_1,\lambda_2)
  = V(\lambda_1,\lambda_2)
  = W(\lambda_1,\lambda_2)
  = -{3\over 4} U(\lambda_1,\lambda_2) \,,
\label{chain}
\end{equation}
where $\lambda_1$ and $\lambda_2$ are to satisfy the relation
(\ref{specc}). It can be proved that to determine all solutions of
this set it is sufficient to solve the equation
\begin{equation}
     V(\lambda_1,\lambda_2)=W(\lambda_1,\lambda_2) \, ,
\end{equation}
which has the explicit form
\begin{equation}
     3(\lambda_2+1)(1-\zeta^4)
   = 7(\lambda_2-3)(\zeta^4-6\zeta^2+1) \, ,
\label{expl}
\end{equation}
and then to verify that the other equalities in (\ref{chain})
hold for these solutions.

To solve Eq.~(\ref{expl}), consider the first branch of Eq.~(\ref{specc}),
\begin{equation}
    \lambda_1^2=2\lambda_2^2+6\lambda_2 \, ,
\label{branch1}
\end{equation}
for which
\begin{equation}
    \zeta = \tan \beta = \frac{2\lambda_2}{\lambda_1} \, .
\end{equation}
Then Eq.~(\ref{expl}) is recast to
\begin{equation}
     (\lambda_2-3) (5\lambda_2^2+24\lambda_2-9) = 0 \,.
\end{equation}
One root of this equation is obviously $\lambda_2=3$, which
corresponds to $\lambda_1=6$.  Another pair of roots is given by
\begin{equation}
     \lambda_2
   = {3\over 5} \left(\pm\sqrt{21}-4\right)\approx 0.350,-5.150 \,,
\label{rts1}
\end{equation}
yielding respectively
\begin{equation}
     \lambda_1
   ={3\over 5}\sqrt{2(17\mp 3\sqrt{21})} \approx 1.530, 4.705 \,.
\end{equation}

Now consider the second branch of Eq.~(\ref{specc}),
\begin{equation}
     \lambda_2 = {\lambda_1^2\over 2} + 1 \,,
\label{branch2}
\end{equation}
with
\begin{equation}
     \zeta= -{2\over\lambda_1}\,.
\end{equation}
In this case, Eq.~(\ref{expl}) becomes
\begin{equation}
     (\lambda_2-3)(\lambda_2^2-26\lambda_2+29) = 0 \,.
\end{equation}
Its roots are $(\lambda_2,\lambda_2) = (3,2)$ and
\begin{eqnarray}
      \lambda_2
  &=& 13\mp 2\sqrt{35} \approx 1.168,  24.83 \,, \nonumber \\
      \lambda_1
  &=& 2\sqrt{6\mp \sqrt{35}} \approx 0.579, 6.904 \,.
\label{rts2}
\end{eqnarray}
Insertion of these solutions into other (\ref{chain}) results in a
chain of identities, as required.

\section{Phases of $^3$P$_2$--$^3$F$_2$ Pairing}

The set of solutions of the $^3$P$_2$--$^3$F$_2$ pairing problem
revealed by the above analysis is depicted in Fig.~1 and
cataloged in Table~1.

Figure 1 refers to the multicomponent solutions and represents
them by their coordinates $(\lambda_1,\lambda_2)$ in the two-dimensional
parameter space.  Only the right half of the $\lambda_1-\lambda_2$ plane
is plotted, because the pairing energies are independent of the sign
of $\lambda_1$.  The most important message of this figure is that
(with the exception previously noted) the solution curves that represent
parametrically degenerate solutions of the $^3$P$_2$ problem shrink to
discrete points as the degeneracy is lifted by the perturbation that
admixes the $^3$F$_2$ channel.

\newpage
%fig. 1 here
\begin{figure}
\begin{center}
\mbox{\psfig{figure=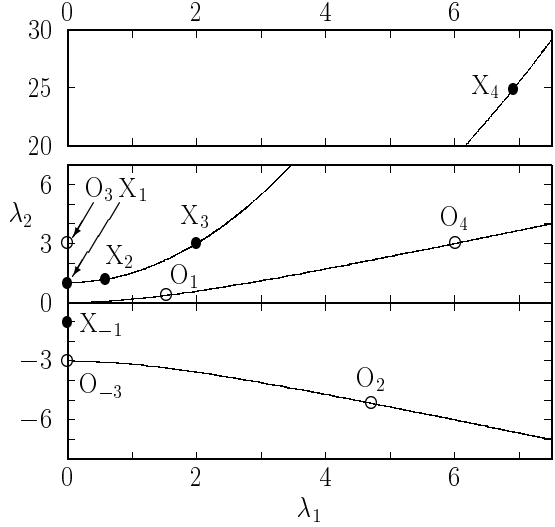,width=7.5cm,height=7cm}}
\end{center}
\caption{Parameter sets $(\lambda_1,\lambda_2)$ defining the
multicomponent solutions of the $^3$P$_2$--$^3$F$_2$ pairing problem.
Solution curves or points of the pure, uncoupled $^3$P$_2$ problem are
identified with an open circle [or a filled circle] according as their
order parameters are nodeless [or display nodes].}
\end{figure}

\begin{table*}
\caption{Identification of the thirteen solutions (or superfluid
phases) of the $^3$P$_2$ pairing model, in terms of the parameters
$\lambda_1$ and $\lambda_2$ defining their magnetic content,
and in terms of their nodeless (states labeled $\rm O$) or nodal
(states labeled $\rm X$) character.}
\begin{center}
\begin{tabular}{|c||c|c|}
\hline
\quad Phase \quad& $\lambda_1$ & $\lambda_2$  \\
\hline
\hline
${\rm O}_{M=0}$ & \multicolumn{2}{|c|} {$M=0$}  \\
\hline
${\rm X}_{M=1}$ & \multicolumn{2}{|c|} {$M=1$}  \\
\hline
${\rm X}_{M=2}$ & \multicolumn{2}{|c|} {$M=2$} \\
\hline
\hline
${\rm O}_3$ & 0 & 3             \\
\hline
${\rm O}_{-3}$ & 0 & $-3$          \\
\hline
${\rm X}_{1}$ & 0 & 1           \\
\hline
${\rm X}_{-1}$ & 0 & $-1$           \\
\hline
\hline
${\rm O}_1$ &
${3\over 5}\sqrt{2(17{-}3\sqrt{21})}$ &
${3\over 5}(\sqrt{21}{-}4)$ \\
\hline
${\rm O}_2$ &
\quad ${3\over 5}\sqrt{2(17{+}3\sqrt{21})}$ \quad &
\quad $-{3\over 5}(\sqrt{21}{+}4)$ \quad \\
\hline
${\rm O}_4$ & 6 & 3           \\
\hline
${\rm X}_2$ &
$2\sqrt{6{-}\sqrt{35}}$ & $13{-}2\sqrt{35}$ \\
\hline
${\rm X}_3$ & 2 & 3          \\
\hline
${\rm X}_4$  &
   $2\sqrt{6{+}\sqrt{35}}$ &
         $13{+}2\sqrt{35}$
                                \\
\hline
\end{tabular}
\end{center}
\end{table*}

\newpage

The solutions (which correspond to pairing states and ultimately to
superfluid phases) divide into two categories: those whose order
parameters contain nodes and those whose order parameters are
nodeless.  It is convenient to identify the nodal states with the
symbol $\rm X$ and the nodeless states with the symbol $\rm O$.

First there are the three well-known ``one-component'' solutions,
belonging to magnetic quantum numbers $M=\pm 2$, $M= \pm 1$, and
$M=0$, respectively.  In addition, we have established the existence
of ten multicomponent solutions that mix states with different values
of $|M|$.  These are comprised of five nodeless solutions and five
exhibiting nodes.

The five nodeless multicomponent solutions ${\rm O}_k$ include:
\begin{itemize}
\item[(i)]
Two two-component solutions denoted ${\rm O}_{\pm 3}$, which are {\it
identical} to the ``particular'' solutions found in the pure $^3$P$_2$
case having $\lambda_1=0$ and $\lambda_2 = \pm 3$.
\item[(ii)]
Three three-component solutions associated with the branch (\ref{branch1})
of Eq.~(\ref{specc}).  Two of these, denoted ${\rm O}_1$ and
${\rm O}_4$, derive respectively from the upper root of the pair (\ref{rts1})
and the root $\lambda_2=3$. The third, named ${\rm O}_2$, derives from the
lower root of the pair (\ref{rts1}).
\end{itemize}

The five nodal solutions ${\rm X}_k$ consist of:
\begin{itemize}
\item[(i)]
Two two-component solutions ${\rm X}_{\pm 1}$ that are identical to
the ``particular'' solutions found in the pure $^3$P$_2$ problem
having $\lambda_1 =0$ and $\lambda_2 = \pm 1$.
\item[(ii)]
Three three-component solutions associated with
the branch (\ref{branch2}) of Eq.~(\ref{specc}).  The solutions
${\rm X}_2$ and ${\rm X}_4$ derive respectively from the lower and upper roots
of the pair (\ref{rts2}), while ${\rm X}_3$ derives from the root $\lambda_2=3$.
\end{itemize}

Appealing to continuity in the parameter $\eta$, it may be expected
that the general features of the spectrum of solutions delineated by
the perturbative analysis will persist even when $\eta$ is not
especially small.

\section{Discussion}

In the present article, we have employed the separation method of
ref.~\cite{kkc2} to derive, from the system of nonlinear BCS integral
equations for the components $\Delta^{JM}_L({\bf p})$ of the gap
function, a set of nonlinear algebraic equations for the coefficients
specifying the angular structure of the superfluid states of interest.
Based on this stratagem, we have been able to find all the real solutions
of the $^3{P_2}$--$^3{F_2}$ pairing problem in the regime of vanishingly
small $\eta$.  Their salient feature is a remarkable independence of
the temperature $T$ and of any details of the interaction ${\cal V}$.
In principle, the full collection of complex solutions can also be
obtained along the same lines, but the calculations become much
more cumbersome.

It must be noted that the status of the $^3$P$_2$--$^3$F$_2$ pairing model,
in which contributions from $^3$P$_2\to ^3$P$_0$ or $^3$P$_2\to ^3$P$_1$
transitions are assumed to be unimportant, is somewhat vulnerable.
This assumption should be a safe one in the low-density regions
of a neutron star.  However, it may not hold in the denser core region,
where (i)~the cancellation between different contributions to the\\
P-wave
central component of the scattering amplitude becomes questionable and
(ii)~the amplification of the tensor force due to the pion-exchange
renormalization \cite{mig} becomes pronounced.  These effects, which
increase with density, may overwhelm the spin-orbit component in
the effective pairing interaction, whose strength depends only mildly
on density.  The $^3$P$_0$ pairing channel would then take command and
give rise to a superfluid state analogous to the B-state of liquid
$^3$He.  In the less challenging situation where $^3$P$_2\to ^3$P$_0$
transitions can be described within perturbation theory, their
effects reduce to a renormalization of the matrix elements $s_{MM_1}$,
which acquire an additional term of the form
\begin{equation}
      \delta s_{MM_1}
  \sim \int S^{2M00}_{11}({\bf n}) K_0({\bf n})d{\bf n}
       \int S^{002M_1}_{11}({\bf n}) K_0({\bf n})d{\bf n}  \ .
\label{corr}
\end{equation}
An analogous modification occurs when contributions from
$^3$P$_2\to ^3$P$_1$ transitions are taken into account
perturbatively.

In short, the introduction of new transitions among two-body states results
in the mixing of components having different values of total angular
momentum $J$.  This effect depends very strongly on the proximity to the
critical temperature $T_c$.  It is well known that in the immediate
vicinity of $T_c$, all the BCS equations decouple; hence the various
phases that appear may be classified according to the quantum number $J$.
These considerations apply as well to superfluid $^3$He.  The magnetic dipole
interaction between the $^3$He atoms -- which is responsible for such
a phase separation close to $T_c$ -- triggers the planar state
with $J=1$ \cite{vol,barton}.  This result is unaffected by
strong-coupling corrections.

Let us now turn to a more detailed analysis of the impact of transitions
between pairing states.  We start with the two-component solutions
(\ref{two}), namely $\lambda_1=0,\, \lambda_2=\pm 1,\pm 3$, and recall
that in the $^3{P_2}$--$^3{F_2}$ model, both of the quantities
$B_1(\lambda_1=0)$ and $B_2(\lambda_1=0)$, which behave as
$r_1(\lambda_1=0)$, vanish identically.  As will now be argued, this
property holds in an extended version that allows for transitions
between states with different $J$ values.  Upon setting $\lambda_1=0$
in the equations (\ref{bi}) and (\ref{rhsr}) determining $B_1$
and $B_2$, there apparently remains a single term containing
$r_1(\lambda_1=0,\lambda_2)=\lambda_2\delta s_{12}
(\lambda_1=0,\lambda_2)+\sqrt{6}\delta s_{10}(\lambda_1=0,\lambda_2)$.
But upon closer inspection of the integrands in (\ref{corr}), this term
is also found to vanish.  According to Eq.~(\ref{ko}),
$K_0(\lambda_1=0,{\bf n})$ is an even function of $\cos\varphi$
independently of $\lambda_2$, while $S^{2100}_{11}({\bf n})$,
the explicit form of which is given in the appendix, is an odd
function of $\cos\varphi$.  Nullification of the respective integrals
therefore occurs upon integration over $\varphi$.  Consequently,
the equality $B_1=B_2=0$ applies even when the transitions
$^3$P$_2\to ^3$P$_1$ are incorporated into the description.  It
follows that the spectrum (\ref{two}) of the two-component solutions
of the $^3$P$_2$--$^3$F$_2$ pairing model endures upon inclusion of
the $^3$P$_2\to ^3$P$_0$ and $^3$P$_2\to ^3$P$_1$ transitions.

The opposite situation is encountered when one considers the particular
solution of the $^3$P$_2$--$^3$F$_2$ pairing model defined by the straight
line $\lambda_2=-1$.  The single point $\lambda_1=0,\, \lambda_2=-1$ of
this line does survive as a solution of the pairing problem,
but the other points do not.
To confirm this assertion, we first observe that in Eq.~(\ref{rel1}),
contributions from the corrected values of $r_2$ and $r_0$ are
proportional to
$\lambda_1$ and hence vanish provided $\lambda_1=0$; further,
as seen from Eq.~(\ref{rhsr}), the corrected value of
$r_1(\lambda_1=0)$ is zero, since contributions from the terms
of Eq.~(\ref{corr}) vanish at $\lambda_1=0$.
To reveal other viable points on the straight line, one looks
for the zeroes of the terms in the l.h.s.\ of Eq.~(\ref{rel1})
introduced by $^3{P_2}\to {^3{P_0}}$ and $^3{P_2}\to {^3{P_1}}$
transitions. Direct calculation shows that the term corresponding to the
$^3{P_2}\to {^3{P_1}}$ transition vanishes identically on the
line $\lambda_2=-1$, while that for the $^3{P_2}\to {^3{P_0}}$
transition is proportional to
$\lambda_1\,J_0^2$, where the integral $J_0$ is given by
Eq.~(\ref{intk}).  It can be checked that this expression has
no new zeroes on the straight line $\lambda_2=-1$ in addition to
the point $\lambda_1=0$.

Close inspection reveals
that the revised locations of the other solutions found above
depend on the matrix elements of the interaction ${\cal V}$;
if the ratio $\eta_P/\eta$ is small then the shifts from the old
positions are also small.  Naturally it is of interest to trace the
trajectories of solutions as this ratio is varied, thereby exploring
the phenomenon of triplet pairing in the continuum of cases from
that of dense neutron matter to that of superfluid $^3$He.  This
investigation will be a subject of future work.

Our method is to be compared with Ginzburg-Landau (GL) theory, which is
generally regarded as the standard technique for mapping the spectrum
of the phases of systems with triplet pairing.  In the GL method,
the search for diverse phases is based on the construction of a
suitable free energy functional up to terms of fourth (or even
sixth) power in the gap value $\Delta$.  This approach allows one
to simultaneously evaluate the splitting between the different
phases and efficiently determine the phase diagram.  Another
advantage of the GL procedure resides in the facility of including
strong-coupling corrections \cite{ss} arising from the dependence of
the effective interaction ${\cal V}$ on the gap value, an effect that
becomes important close to the critical temperature $T_c$.  Unfortunately,
the GL method fails when the temperature $T$ is significantly different
from $T_c$.  Its application to the phase-spectrum problem makes a
sense only if the phase structure of superfluid neutron matter is
independent of $T$.  However, special conditions must be met for this
to be true in the face of the explicit appearance of the factor
$\tanh (E/2T)$ in the set of BCS equations, and the explication
of these conditions is impossible within the GL approach itself.  Our
method is free of these shortcomings.  It is equally reliable close to
$T_c$ and at $T=0$.  Moreover, the incorporation of strong-coupling
corrections reduces to the insertion of new terms in ${\cal V}$ that depend
on $\Delta$ itself; if this dependence is appropriately specified,
no further hurdles must be overcome to fully elucidate the triplet
superfluid phase diagram.

In a sequel to this paper, we shall report the findings of a quantitative
treatment of the transitions between the phases arising in the
$^3$P$_2$--$^3$F$_2$ model and construct the corresponding superfluid
phase diagram of dense neutron matter.  This task requires significant
numerical effort beyond the analytic developments of the present
work.

\vskip 0.5 cm {\bf Acknowledgements} \vskip 0.5 cm

This research was supported in part by the U.~S. National Science
Foundation under Grant No. PHY-0140316, by the McDonnell Center
for the Space Sciences at Washington University, and by
Grant No.~00-15-96590 from the Russian Foundation for Basic
Research (VAK and MVZ).  JWC also acknowledges support received
from Funda{\c c}\~ao Luso-Americana para o Desenvolvimento (FLAD)
and from Funda{\c c}\~ao para a Ci\^encia e a Technologia (FCT) for
his participation in Madeira Math Encounters XXIII at
the University of Madiera, where some of this work was done.
He thanks Professor Ludwig Streit and his colleagues for
the generous hospitality extended by the Centro de Ci\^encias
Mathem\'aticas.

\appendix\section{Explicit Formulas for Matrix Elements}

To find analytic solutions of the system (\ref{bc}),
explicit formulas for the set of matrix elements
\begin{eqnarray}
   &&  S^{JMJ_1M_1}_{LL_1}({\bf n})\nonumber\\
   &=& (-1)^{M+1+(L-L_1+K)/2}
       {1\over 4\pi}\sum\limits_{K\kappa}
       \sqrt{(2K+1)(2J+1)(2J_1+1)(2L+1)(2L_1+1)}\nonumber
\end{eqnarray}
\begin{equation}
\label{s13}
\times \left(\begin{array}{ccc}
  K & L & L_1\\
  0 & 0 & 0
\end{array}\right)
\left(\begin{array}{ccc}
  K & J_1 & J\\
  \kappa & M_1 & -M
\end{array}\right)
\left\{\begin{array}{ccc}
  K & J_1 & J\\
  1 & L & L_1
\end{array}\right\}
(-1)^{(\kappa+|\kappa|)/2} {\cal P}_{K\kappa}(\vartheta,\varphi)
e^{-i\kappa\varphi}\,.
\end{equation}
are needed, where the ${\cal P}_{K\kappa}(\vartheta,\varphi)$
are associated Legendre polynomials.
Lengthy algebra involving $3j$- and $6j$-symbols yields
%(A2)
\begin{eqnarray}
     \left(S^{2020}_{13}+S^{2020}_{31}\right)
 &=& {1\over 7\pi} \, \sqrt{3 \over 2}
     \left(-{105 \over 4} \cos^4\vartheta+21\cos^2\vartheta
           -{7\over 4} \right)\,,\\
%(A3)
     \left(S^{2021}_{13}+S^{2021}_{31}\right)
 &=& {1\over 7\pi} \,
     \left( 12\sin\vartheta \cos\vartheta
     -{105\over 4}\sin\vartheta\cos^3\vartheta\right) e^{i\varphi}\,,\\
%(A4)
      \left(S^{202,-1}_{13}+S^{202,-1}_{31}\right)
 &=& -{1\over 7\pi} \,
      \left( 12\sin\vartheta \cos\vartheta
      -{105\over 4}\sin\vartheta\cos^3\vartheta\right) e^{-i\varphi}\,,\\
%(A5)
     \left(S^{2022}_{13}+S^{2022}_{31}\right)
 &=& {1\over 7\pi} \,
     \left( {21\over 8}\sin^2\vartheta
           -{105\over 8}\sin^2\vartheta\cos^2\vartheta\right)
     e^{2i\varphi}\,,\\
%(A6)
     \left(S^{202,-2}_{13}+S^{202,-2}_{31}\right)
 &=& {1\over 7\pi} \,
     \left( {21\over 8}\sin^2\vartheta
     -{105\over 8}\sin^2\vartheta\cos^2\vartheta\right)
      e^{-2i\varphi}\,,\\
%(A7)
     \left(S^{2120}_{13}+S^{2120}_{31}\right)
 &=& {1\over 7\pi} \,
     \left( 12\sin\vartheta\cos\vartheta
     -{105\over 4}\sin\vartheta\cos^3\vartheta\right)
      e^{-i\varphi}\,,\\
%(A8)
      \left(S^{2121}_{13}+S^{2121}_{31}\right)
 &=& {1\over 7\pi} \, \sqrt{3\over 2}
     \left( {35 \over 2} \cos^4\vartheta
     - {63\over 4}\cos^2\vartheta + {7\over 4} \right)\,,\\
%(A9)
      \left(S^{212,-1}_{13}+S^{212,-1}_{31}\right)
 &=& -{1\over 7\pi} \, \sqrt{3 \over 2}
      \left( {35 \over 2} \cos^4\vartheta
      - {77\over 4}\cos^2\vartheta + {7\over 4}\right)e^{-2i\varphi}\\
%(A10)
     \left(S^{2122}_{13}+S^{2122}_{31}\right)
 &=& {1\over 7\pi} \, \sqrt{3 \over 2}
     \left( -{9\over 4} \sin\vartheta \cos\vartheta
          +{35\over 4}\sin\vartheta\cos^2\vartheta \right)
            e^{i\varphi}\,,\\
%(A11)
      \left(S^{212,-2}_{13}+S^{212,-2}_{31}\right)
  &=& -{1\over 7\pi} \, {35 \over 4}\sqrt{3 \over 2}
      \sin^3\vartheta \cos\vartheta e^{-3i\varphi}\,,\\
%(A12)
      \left(S^{2220}_{13}+S^{2220}_{31}\right)
  &=& {1\over 7\pi} \,
      \left( {21\over 8}\sin^2\vartheta-{105\over 8}
      \sin^2\vartheta\cos^2\vartheta \right) e^{-2i\varphi}\,,\\
%(A13)
      \left(S^{2221}_{13}+S^{2221}_{31}\right)
  &=& {1\over 7\pi} \, \sqrt{35 \over 2}
      \left( -{9\over 4} \sin\vartheta \cos\vartheta+{35\over 4}
           \sin\vartheta\cos^2\vartheta \right)
      e^{-i\varphi},\\
%(A14)
      \left(S^{222,-1}_{13}+S^{222,-1}_{31}\right)
  &=& {1\over 7\pi} \, {35 \over 4 }\sqrt{3 \over 2}
      \sin^3\vartheta \cos\vartheta e^{-3i\varphi}\,,\\
%(A15)
      \left(S^{2222}_{13}+S^{2222}_{31}\right)
  &=& -{1\over 7\pi} \, \sqrt{3 \over 2}
       \left( {7\over 8} + {21\over 4}\cos^2\vartheta
       - {35\over 8} \cos^4\vartheta \right)\,,\\
%(A16)
       \left(S^{222,-2}_{13}+S^{222,-2}_{31}\right)
  &=& -{1\over 7\pi} \, {35 \over 8}\sqrt{3 \over 2}
        \sin^4\vartheta \cos\vartheta e^{-4i\varphi}\,.
\end{eqnarray}

The combinations entering Eqs.~(\ref{rhsr}) for $r_M$ are
%(A17)
\begin{eqnarray}
      S_{00}(\vartheta,\varphi)
  &=& \mbox{Re}\,S^{2020}_{13} +\mbox{Re}\,S^{2020}_{31}\nonumber\\
  &=& \sqrt{3\over 2} \left( -{105 \over 4} \cos^4\vartheta
       +21\cos^2\vartheta-{7\over 4} \right)\, ,
\end{eqnarray}
%(A18)
\begin{eqnarray}
     S_{01}(\vartheta,\varphi)
 &=& \mbox{Re}\,S^{2021}_{13}+\mbox{Re}\,S^{2021}_{31}
     -\mbox{Re}\,S^{202,-1}_{13}-\mbox{Re}\,S^{202,-1}_{31}\nonumber\\
 &=& \left( 24 \cos\vartheta\sin\vartheta
     -{105\over 2}\sin\vartheta\cos^3\vartheta \right) \cos\varphi\, ,
\end{eqnarray}

%(A19)
\begin{eqnarray}
      S_{02}(\vartheta,\varphi)
  &=& \mbox{Re}\,S^{2022}_{13}+\mbox{Re}\,S^{2022}_{31}
      +\mbox{Re}\,S^{202,-2}_{13}+\mbox{Re}\,S^{202,-2}_{31}\nonumber\\
  &=& \left( {21\over 4} \sin^2\vartheta
            -{105\over 2}\sin^2\vartheta\cos^2\vartheta \right)
             \cos 2\varphi\, ,
\end{eqnarray}
%(A20)
\begin{eqnarray}
      S_{10}(\vartheta,\varphi)
  &=& \mbox{Re}\, S^{2120}_{13}+\mbox{Re}\,S^{2120}_{31}\nonumber\\
  &=& \left( 12 \cos\vartheta\sin\vartheta
      -{105\over 4}\sin\vartheta\cos^3\vartheta \right) \cos\varphi\, ,
\end{eqnarray}
%(A21)
\begin{eqnarray}
      S_{11}(\vartheta,\varphi)
  &=& \mbox{Re}\,S^{2121}_{13}+\mbox{Re}\,S^{2121}_{31}
      -\mbox{Re}\,S^{212,-1}_{13}-\mbox{Re}\,S^{212,-1}_{31}\nonumber\\
  &=& \sqrt{3\over 2}
      \left( {35 \over 2}\cos^4\vartheta
            -{63\over 4}\cos^2\vartheta + {7\over 4} \right)\nonumber\\
  &\quad&~ + \sqrt{3\over 2}
           \left( {35\over 2}\cos^4\vartheta
                 -{77\over 4}\cos^2\vartheta
                 +{7\over 4} \right) \cos 2\varphi\, ,
\end{eqnarray}
%(A22)
\begin{eqnarray}
      S_{12}(\vartheta,\varphi)
  &=& \mbox{Re}\,S^{2122}_{13}+\mbox{Re}\,S^{2122}_{31}
     +\mbox{Re}\,S^{212,-2}_{13}+\mbox{Re}\,S^{212,-2}_{31}\nonumber\\
  &=& \sqrt{3\over 2} \left( -{9 \over 4} \sin\vartheta\cos\vartheta
       + {35\over 4}\sin\vartheta\cos^3\vartheta\right)\cos\varphi\nonumber\\
  &\quad&~ -{35 \over 4} \sqrt{3\over 2}
          \cos\vartheta\sin^3\vartheta\cos 3\varphi\, ,
\end{eqnarray}
%(A23)
\begin{eqnarray}
      S_{20}(\vartheta,\varphi)
  &=& \mbox{Re}\,S^{2220}_{13}+\mbox{Re}\,S^{2220}_{31}\nonumber\\
  &=& \left( {21\over 8} \sin^2\vartheta
      -{105\over 8}\sin^2\vartheta\cos^2\vartheta \right) \cos 2\varphi\, ,
\end{eqnarray}
%(A24)
\begin{eqnarray}
     S_{21}(\vartheta,\varphi)
 &=& \mbox{Re}\,S^{2221}_{13}+\mbox{Re}\,S^{2221}_{31}
     -\mbox{Re}\,S^{222,-1}_{13}-\mbox{Re}\,S^{222,-1}_{31}\nonumber\\
 &=& \sqrt{3\over 2}
     \biggl( -{9\over 4} \sin\vartheta \cos\vartheta\nonumber\\
 &\quad&~ +{35\over 4}\sin\vartheta\cos^2\vartheta \biggr)
         \cos\varphi - {35 \over 4} \sqrt{3\over 2}
         \sin^3\vartheta\cos\vartheta \cos 3\varphi\, ,
\end{eqnarray}
%(A25)
\begin{eqnarray}
      S_{22}(\vartheta,\varphi)
  &=& \mbox{Re}\,S^{2222}_{13}+\mbox{Re}\,S^{2222}_{31}
      +\mbox{Re}\,S^{222,-2}_{13}+\mbox{Re}\,S^{222,-2}_{31}\nonumber\\
  &=& -\sqrt{3\over 2} \left( {35 \over 8} \cos^4\vartheta
      - {21\over 4}\cos^2\vartheta + {7\over 8} \right)\nonumber\\
  &\quad&~ -{35 \over 8} \sqrt{3\over 2} \sin^4\vartheta \cos 4\varphi\, .
\end{eqnarray}
Further calculations are conveniently carried out in terms of the
harmonic variables $x=\sin\vartheta\cos\varphi$,
$y=\sin\vartheta\sin\varphi$, and $z=\cos\vartheta$,
which satisfy $x^2+y^2+z^2=1$. One has
\begin{eqnarray}
\label{relat}
     \sin^2\vartheta\cos 2\varphi
 &=& 2x^2+z^2-1 \,,
     \quad \sin^2\vartheta\cos^2\vartheta\cos 2\varphi
            =2x^2z^2+z^4-z^2 \,,\nonumber\\
     \sin^4\vartheta\cos 4\varphi
 &=& 8x^4-8x^2+8x^2z^2+1-2z^2+z^4 \,,
     \quad \sin\vartheta\cos^3\vartheta\cos\varphi=xz^3 \, ,\nonumber\\
      \cos\vartheta\sin^3\vartheta\cos 3\varphi
 &=& 4zx^3-zx+3z^3x \,,
     \quad \sin\vartheta\cos\vartheta\cos\varphi=xz \, .
%\label{relat}
\end{eqnarray}
Accounting for these relations, one finds
%(A27)
\begin{eqnarray}
      S_{00}(x,z)
  &=& {7\over 4}\sqrt{3\over 2}
      \left( -15z^4+12z^2-1 \right) \, ,
        \quad S_{01}(x,z)
              = {3\over 2}\left( 16xz-35xz^3 \right) \, ,\nonumber\\
       S_{02}(x,z)
  &=& {21\over 4}
      \left( 2x^2+6z^2-10x^2z^2-5z^4-1 \right) \, ,\nonumber\\
      S_{10}(x,z)
  &=& {1\over 4} \left( 48xz-105xz^3 \right) \, ,
      \quad S_{11} = {7\over 2}\sqrt{3\over 2}
            \left( z^2+x^2-10x^2z^2 \right) \, ,\nonumber\\
      S_{12}(x,z)
  &=& {1\over 2}\sqrt{3\over 2}
      \left( 48xz-35xz^3-70x^3z \right) \, ,\nonumber\\
      S_{20}(x,z)
  &=& {21\over 8}
      \left( 2x^2+6z^2-10x^2z^2-5z^4-1 \right) \, ,\nonumber\\
      S_{21}(x,z)
  &=& {1\over 2}\sqrt{3\over 2}
      \left( 48xz-35xz^3-70x^3z \right) \, ,\nonumber\\
      S_{22}(x,z)
  &=& -{7\over 4}\sqrt{3\over 2}
      \left( 5z^4-8z^2+20x^4-20x^2+20x^2z^2+3 \right) \, .
\label{sss}
\end{eqnarray}
We may now evaluate the combinations $R_M(x,z)$ entering Eq.~(\ref{psi12}):
%(A28)
\begin{eqnarray}
           R_0(x,z)
 &\equiv& {\lambda_2\over \sqrt{6}} S_{02}(x,z)
          +{\lambda_1\over \sqrt{6}} S_{01}(x,z)
          + S_{00}(x,z) \nonumber \\
 &=& {\sqrt{3}\over 4\sqrt{2}}
     \Bigl\{ 7\bigl[ 2\lambda_2 x^2 + 6(\lambda_2+2)z^2
     - 10\lambda_2 x^2 z^2\nonumber\\
 &\quad&~ -5(\lambda_2+3)z^4-\lambda_2-1\bigr]
         +2\lambda_1(16xz-35xz^3) \Bigr\} \,,
\label{rr0}
\end{eqnarray}
%(A29)
\begin{eqnarray}
          R_1(x,z)
 &\equiv& \lambda_2 S_{12}(x,z) + \lambda_1 S_{11}(x,z)
          + \sqrt{6} S_{10}(x,z) \nonumber \\
 &=& {\sqrt{3}\over 2\sqrt{2}}
     \Bigl\{ \bigl[48(\lambda_2+1)xz - 35(\lambda_2+3)xz^3
            - 70\lambda_2 x^3 z \bigr] \nonumber \\
 &\quad&~ +7\lambda_1(z^2+x^2-10x^2z^2) \Bigr\} \,,
\label{rr1}
\end{eqnarray}
%(A30)
\begin{eqnarray}
            R_2(x,z)
   &\equiv& \lambda_2 S_{22}(x,z) + \lambda_1 S_{21}(x,z)
            + \sqrt{6} S_{20}(x,z) \nonumber \\
   &=& {\sqrt{3}\over 4\sqrt{2}}
       \Bigl\{ 7\bigl[ -20\lambda_2 x^4
              + 2(10\lambda_2+3)x^2\nonumber\\
   &\quad&~ -10(2\lambda_2+3) x^2 z^2 + 2(4\lambda_2+9)z^2
           - 5(\lambda_2+3)z^4-3\lambda_2-3 \bigr]\nonumber\\
   &\quad&~ +2\lambda_1(48xz-35xz^3-70x^3z) \Bigr\} \,.
\label{rr2}
\end{eqnarray}
Substitution of Eq.~(\ref{sss}) into Eqs.~(\ref{rr0})-(\ref{rr2})
leads to the results
%(A31)
\begin{eqnarray}
      R_0(x,z)
 &=& {\sqrt{3}\over 4\sqrt{2}}
     \Bigl\{ 7\bigl[ 2\lambda_2 x^2
            +6(\lambda_2+2)z^2 - 10\lambda_2 x^2 z^2\nonumber\\
 &\quad&~ -5(\lambda_2+3)z^4-\lambda_2-1\bigr]
         +2\lambda_1(16xz-35xz^3) \Bigr\} \,,
\end{eqnarray}
%(A32)
\begin{eqnarray}
     R_1(x,z)
 &=& {\sqrt{3}\over 2\sqrt{2}}
     \Bigl\{ \bigl[48(\lambda_2+1)xz
            - 35(\lambda_2+3)xz^3 - 70\lambda_2 x^3 z \bigr] \nonumber \\
 &\quad&~ +7\lambda_1(z^2+x^2-10x^2z^2) \Bigr\} \,,
\end{eqnarray}
%(A33)
\begin{eqnarray}
      R_2(x,z)
  &=& {\sqrt{3}\over 4\sqrt{2}}
      \Bigl\{ 7\bigl[ -20\lambda_2 x^4 + 2(10\lambda_2+3)x^2
      - 10(2\lambda_2+3) x^2 z^2\nonumber\\
  &\quad&~ + 2(4\lambda_2+9)z^2
          -5(\lambda_2+3)z^4 -3\lambda_2-3\bigr]\nonumber\\
  &\quad&~ +2\lambda_1 (48xz-35xz^3-70 x^3z)\Bigr\}.
\end{eqnarray}

\end{document}